\documentclass[pra,twocolumn,tightenlines,showpacs,floatfix,nofootinbib]{revtex4}
\usepackage[active]{srcltx} 
\usepackage{bm,dcolumn,amsmath,graphicx}

\begin{document}
\title{Laser cooling and trapping of potassium at magic wavelengths}

\author{M. S. Safronova$^{1,2}$}
\author{U. I. Safronova$^{3,4}$}
\author{Charles W. Clark$^{2}$}

\affiliation {$^1$Department of Physics and Astronomy, 217 Sharp Lab, University of Delaware, Newark, Delaware 19716
 \\$^2$Joint Quantum Institute, National Institute of Standards and Technology
and the University of Maryland, Gaithersburg, Maryland 20899-8410, USA\\
 $^3$Physics Department, University of Nevada,
Reno, Nevada 89557
\\$^4$Department of Physics,  University of Notre Dame,
Notre Dame, Indiana 46556}

\begin{abstract}

We carry out a systematic study of the static and dynamic polarizabilities of the potassium atom using a
first-principles high-precision relativistic all-order
 method in which all single, double, and partial triple excitations
of the Dirac-Fock wave functions
 are included to all orders of perturbation theory.
 Recommended values
 are provided for a large number of
electric-dipole matrix elements. Static polarizabilities of the $4s$, $4p_j$, $5s$, $5p_j$, and $3d_j$ states are
compared with other theory and experiment where available. We use the results of the polarizability calculations  to
identify magic wavelengths  for the
 $4s-np$
transitions for $n = 4, 5$, {\em i.e.} those wavelengths for which the two levels have the same ac Stark shifts.  These
facilitate state-insensitive optical cooling and trapping. The magic wavelengths for the $4s-5p$ transitions
are of particular interest for attaining a quantum gas of potassium at high phase-space density.
We find 20 such wavelengths in the technically interest region of $1050-1130$~nm. 
  Uncertainties of all
recommended values are estimated.
\end{abstract}
\maketitle

\section{Introduction}

Due to their applications in ultra-precise atomic clocks, degenerate quantum gases and quantum information, the magic
wavelengths of atoms have become a subject of great interest in both experiments
\cite{YiMejMcF11,MckJerFin11,LunSchPor10,McFMejYi10} and theory
\cite{Der10,KatHasIli09,DzuFlaLev11,ZhaRobSaf11,DamSWil12,GenJiaRun11,SafJiaKoz10,YeWan08,GuoWanYe10,AroSafCla07,YunXiaJin06}.
The energy levels of atoms trapped in a light field are shifted by an amount that is proportional to their
frequency-dependent polarizability, so the difference in the energies of any two levels depends upon the trapping
field.  This difference is often called the ``ac Stark shift''.

The idea of a ``magic'' wavelength, $\lambda_{\rm{magic}}$, at which
there is no relative shift of a given pair of energy levels, was
first proposed in Refs.~\cite{KatIdoKuw99,YeVerKim99} in the context
of optical atomic clocks. An atom confined in a trap constructed of
light with a magic wavelength for the clock transition will, to
lowest order, have the same transition energy as it does in free
space.

 This simple idea has a number of other applications.  A problem arises in cooling and trapping schemes, where the ac Stark shift of the cooling or trapping transition may lead to heating.
Recent experiments in $^6$Li \cite{DuaHarHit11} and $^{40}$K \cite{MckJerFin11} degenerate quantum gases in optical
traps demonstrated temperature reductions by a factor of about five and phase-space density increases by at least a
factor of ten by laser cooling using ultraviolet (UV) transitions ($2s-3p$ and $4s-5p$, respectively) compared to
conventional cooling with the visible or infrared $D_1$ and $D_2$ transitions. However, the ac Stark shifts due to trap
light must be nearly the same for both levels in the transition to allow for efficient and uniform cooling
\cite{DuaHarHit11}. This is accomplished by building the optical trap using light with the magic
   wavelength for the corresponding UV transitions.
  The use of the magic wavelengths is also advantageous for trapping and controlling atoms in high-Q cavities in
 the strong coupling regime, so as to minimize decoherence
in quantum computation and communication protocols \cite{MckBucBoo03}, and in the implementation of the Rydberg gate
for quantum computing with neutral atoms~\cite{SafWilCla03,Saffman}.

Variations on the magic wavelength idea include the use of multiple light fields to attain ac Stark shift targets
~\cite{AroSafCla10} or to maximize differential response between different atomic species -- for example, the
``tune-out'' wavelengths that trap one species but not another~\cite{LebThy07,AroSafCla11}.  Design and evaluation of
all these applications requires accurate data on atomic frequency-dependent polarizabilities. One goal of our present
work is to provide a list of all magic wavelengths for potassium UV $4s-5p_j$ transitions in regions that are
convenient for laser cooling of ultracold gases to high phase-space densities.  For example, in 2011, low-temperature
high-density magneto-optical trapping of potassium using the open $4s_{1/2}\rightarrow 5p_{3/2}$ transition at 405~nm was performed
by McKay {\it et al.\/} \cite{MckJerFin11}. Fermionic $^{40}$K was captured using a magneto-optical trap (MOT) on the
closed $4s\rightarrow 4p_{3/2}$ transition at 767~nm and then transferred, with high efficiency, to a MOT on the open
$4s\rightarrow  5p_{3/2}$ transition at 405~nm. Because the $5p_{3/2}$ state has a smaller linewidth than the
$4p_{3/2}$ state, the Doppler limit is reduced from 145$\mu$K  to 24$\mu$K, and
temperatures as low as 63(6)$\mu$K were observed.

In this paper we provide a list of
magic wavelengths for the $4s-4p$ and $4s-5p$ transitions, calculate
dc and ac polarizabilities for several low-lying states, and provide
recommended values for a number of relevant electric-dipole
transitions which are of interest to applications such as those
described above. Where possible, we compare our results with available experimental \cite{HolRevLon10} and
high-precision theoretical values \cite{SafSaf08}.

Some of the calculations reported here required evaluation of the
electric-dipole matrix elements for very highly excited states, such
as $14s$. These states are needed since the ac polarizabilities for
the magic wavelengths of particular experimental interest (around
1050~nm) are dominated by the $5p-nl$ transitions with $n=12-14$.
Such states were previously beyond the capabilities of the all-order
method used here due to the large spatial extent of the orbitals. In
this work, we resolved the numerical problems associated with such
calculations and successfully demonstrated the stability of our new
approach.

We begin with a brief review of recent research on the applications of magic wavelength concepts in Section~\ref{seca}.
The calculation of electric-dipole
 matrix elements, static and dynamic polarizabilities as well as their
uncertainties is discussed in Section~\ref{sec1}.  The magic
wavelengths are discussed in Section~\ref{sec3}.

\section{Review of magic wavelength studies}
\label{seca} Up to the present, most work on magic wavelengths has
been done on group I and group II atoms, which are the species most
easily cooled and trapped by optical methods.  We summarize these in
turn. The following examples are representative of significant
applications of the magic wavelength concept, however, these examples
are not intended to constitute an exhaustive review.
\subsection{Group I}
\label{secaI} The cancellation of the differential ac Stark shift of
the microwave hyperfine clock transition in trapped $^{87}$Rb atoms
was demonstrated in \cite{LunSchPor10}. The technique  had
implications for experiments involving the precise control of
optically trapped neutral atoms, but
 the cancellation comes at the expense of a small magnetic-field sensitivity.
``Doubly magic'' conditions in magic-wavelength trapping of ultracold
alkali-metal atoms were investigated by Derevianko \cite{Der10}. This
work demonstrated that the microwave transitions in alkali-metal
atoms may be indeed made impervious to both trapping laser intensity
and fluctuations of magnetic fields.

The issue of the mismatch of the polarizabilities of the ground and excited states has also arisen in  the Rydberg gate
approach to quantum information processing \cite{BreDeuJes00,SafWalMol10}, in which the qubit is based on two ground hyperfine states of neutral atoms
confined in an optical lattice. An atom in a Rydberg state will, in general, move in a different optical lattice
potential than that experienced by the ground state. Therefore, the vibrational state of the atom in the lattice may
change after the gate operation is completed, leading to decoherence due to motional heating. This problem may be
resolved by the use of magic wavelengths~\cite{SafWilCla03,Saffman}.
 Use of magic wavelengths in optical traps for
Rydberg atoms was also discussed in \cite{ZhaRobSaf11}, where
 three designs for blue-detuned dipole traps were presented.

\begin{table*}[ht]
 \caption{Absolute values of the reduced electric-dipole matrix
elements in K and their uncertainties in a.u.. The present all-order values are given unless noted otherwise.
$^{(a)}$Expt.~\cite{FalSheTie06}, $^{(b)}$determined from Stark shift data in ~\cite{AroSafCla07}. The uncertainties are estimated where possible (see text). } \label{tab1}
\begin{ruledtabular}
\begin{tabular}{lrlrlrlr}
\multicolumn{1}{c}{Transition}& \multicolumn{1}{c}{Value}&
\multicolumn{1}{c}{Transition}& \multicolumn{1}{c}{Value}&
\multicolumn{1}{c}{Transition}& \multicolumn{1}{c}{Value}&
\multicolumn{1}{c}{Transition}&
\multicolumn{1}{c}{Value}\\
\hline
$ 4s       -  4p_{1/2}$&4.106(4)$^{(a)}$ &$ 5s       -  4p_{1/2}$&        3.885(8) &$ 4s       -  4p_{3/2}$&       5.807(7)$^{(a)}$ &$ 5s       -  4p_{3/2}$&         5.54(1)\\
$ 4s       -  5p_{1/2}$&       0.2755   &$ 5s       -  5p_{1/2}$&         9.49(3) &$ 4s       -  5p_{3/2}$&       0.4060   &$ 5s       -  5p_{3/2}$&        13.40(4)\\
$ 4s       -  6p_{1/2}$&       0.0855   &$ 5s       -  6p_{1/2}$&         0.90(1) &$ 4s       -  6p_{3/2}$&       0.1302   &$ 5s       -  6p_{3/2}$&         1.30(2)\\
$ 4s       -  7p_{1/2}$&       0.0390   &$ 5s       -  7p_{1/2}$&        0.3347   &$ 4s       -  7p_{3/2}$&       0.0614   &$ 5s       -  7p_{3/2}$&        0.4907  \\
$ 4s       -  8p_{1/2}$&       0.0225   &$ 5s       -  8p_{1/2}$&        0.183(3) &$ 4s       -  8p_{3/2}$&       0.0364   &$ 5s       -  8p_{3/2}$&        0.271(4)\\
$ 4s       -  9p_{1/2}$&       0.0147   &$ 5s       -  9p_{1/2}$&        0.120(2) &$ 4s       -  9p_{3/2}$&       0.0244   &$ 5s       -  9p_{3/2}$&        0.178(3)\\
$ 4s       - 10p_{1/2}$&       0.0105   &$ 5s       - 10p_{1/2}$&        0.087(1) &$ 4s       - 10p_{3/2}$&       0.0177   &$ 5s       - 10p_{3/2}$&        0.129(2)\\[0.4pc]
$ 4p_{1/2} -  6s      $&       0.903(4) &$ 5p_{1/2} -  6s      $&         8.79(2)& $ 4p_{3/2} -  6s      $&       1.279(5) &$ 5p_{3/2} -  6s      $&        12.50(2)\\
$ 4p_{1/2} -  7s      $&       0.476(2) &$ 5p_{1/2} -  7s      $&        1.801(8)& $ 4p_{3/2} -  7s      $&       0.673(3) &$ 5p_{3/2} -  7s      $&         2.54(1)\\
$ 4p_{1/2} -  8s      $&       0.314(2) &$ 5p_{1/2} -  8s      $&        0.912(5)& $ 4p_{3/2} -  8s      $&       0.444(2) &$ 5p_{3/2} -  8s      $&        1.287(7)\\
$ 4p_{1/2} -  9s      $&       0.230(1) &$ 5p_{1/2} -  9s      $&        0.592(3)& $ 4p_{3/2} -  9s      $&       0.325(2) &$ 5p_{3/2} -  9s      $&        0.834(4)\\
$ 4p_{1/2} - 10s      $&      0.1791(9) &$ 5p_{1/2} - 10s      $&        0.430(2)& $ 4p_{3/2} - 10s      $&       0.253(1) &$ 5p_{3/2} - 10s      $&        0.607(3)\\
$ 4p_{1/2} - 11s      $&      0.1452(8) &$ 5p_{1/2} - 11s      $&        0.334(2)& $ 4p_{3/2} - 11s      $&       0.205(1) &$ 5p_{3/2} - 11s      $&        0.471(3)\\[0.4pc]
$ 4p_{1/2} -  3d_{3/2}$&        7.979(35)$^{(b)}$ &$ 5p_{1/2} -  3d_{3/2}$&          7.2(1)&  &  &$ 5p_{3/2} -  3d_{3/2}$&         3.19(5)\\
$ 4p_{1/2} -  4d_{3/2}$&      0.1121(8) &$ 5p_{1/2} -  4d_{3/2}$&        17.04(6)&  $ 4p_{3/2} -  4d_{3/2}$&      0.0400(1) &$ 5p_{3/2} -  4d_{3/2}$&         7.64(3)\\
$ 4p_{1/2} -  5d_{3/2}$&       0.333(2) &$ 5p_{1/2} -  5d_{3/2}$&        0.931(4)&  $ 4p_{3/2} -  5d_{3/2}$&       0.155(1) &$ 5p_{3/2} -  5d_{3/2}$&        0.398(2)\\
$ 4p_{1/2} -  6d_{3/2}$&       0.341(3) &$ 5p_{1/2} -  6d_{3/2}$&        0.063(6)&  $ 4p_{3/2} -  6d_{3/2}$&       0.157(1) &$ 5p_{3/2} -  6d_{3/2}$&        0.039(3)\\
$ 4p_{1/2} -  7d_{3/2}$&       0.298(2) &$ 5p_{1/2} -  7d_{3/2}$&        0.219(5)&  $ 4p_{3/2} -  7d_{3/2}$&       0.136(1) &$ 5p_{3/2} -  7d_{3/2}$&        0.105(3)\\
$ 4p_{1/2} -  8d_{3/2}$&       0.254(2) &$ 5p_{1/2} -  8d_{3/2}$&        0.236(4)&  $ 4p_{3/2} -  8d_{3/2}$&       0.116(1) &$ 5p_{3/2} -  8d_{3/2}$&        0.111(2)\\
$ 4p_{1/2} -  9d_{3/2}$&       0.218(2) &$ 5p_{1/2} -  9d_{3/2}$&        0.222(3)&  $ 4p_{3/2} -  9d_{3/2}$&      0.0995(8) &$ 5p_{3/2} -  9d_{3/2}$&        0.103(2)\\[0.4pc]
$ 4p_{3/2} -  3d_{5/2}$&       10.734(47)$^{(b)}$ &$ 5p_{3/2} -  3d_{5/2}$&          9.6(1)& &  &&   \\
$ 4p_{3/2} -  4d_{5/2}$&      0.1170(4) &$ 5p_{3/2} -  4d_{5/2}$&        22.93(8)& $ 3d_{3/2} -  5p_{1/2}$&      7.2(1) &$ 3d_{5/2} -  5p_{3/2}$&      9.6(1)\\
$ 4p_{3/2} -  5d_{5/2}$&       0.467(3) &$ 5p_{3/2} -  5d_{5/2}$&        1.188(7)& $ 3d_{3/2} -  6p_{1/2}$&     1.03(1) &$ 3d_{5/2} -  6p_{3/2}$&     1.39(1)\\
$ 4p_{3/2} -  6d_{5/2}$&       0.471(4) &$ 5p_{3/2} -  6d_{5/2}$&        0.119(8)& $ 3d_{3/2} -  7p_{1/2}$&    0.497(5) &$ 3d_{5/2} -  7p_{3/2}$&    0.673(7)\\
$ 4p_{3/2} -  7d_{5/2}$&       0.409(3) &$ 5p_{3/2} -  7d_{5/2}$&        0.318(7)& $ 3d_{3/2} -  8p_{1/2}$&    0.317(3) &$ 3d_{5/2} -  8p_{3/2}$&    0.428(4)\\
$ 4p_{3/2} -  8d_{5/2}$&       0.349(3) &$ 5p_{3/2} -  8d_{5/2}$&        0.335(6)& $ 3d_{3/2} -  9p_{1/2}$&    0.228(3) &$ 3d_{5/2} -  9p_{3/2}$&    0.308(4)\\
$ 4p_{3/2} -  9d_{5/2}$&       0.299(2) &$ 5p_{3/2} -  9d_{5/2}$&        0.312(4)& $ 3d_{3/2} - 10p_{1/2}$&    0.176(2) &$ 3d_{5/2} - 10p_{3/2}$&    0.238(3)\\[0.4pc]
$ 3d_{3/2} -  4p_{3/2}$&     3.578(16)$^{(b)}$ & $ 3d_{3/2} -  4f_{5/2}$&     12.3(2) & $ 3d_{5/2} -  4f_{7/2}$&     14.6(2)&$ 3d_{5/2} -  4f_{5/2}$&     3.27(4)\\
$ 3d_{3/2} -  5p_{3/2}$&     3.19(5) & $ 3d_{3/2} -  5f_{5/2}$&     4.92(2) & $ 3d_{5/2} -  5f_{7/2}$&     5.88(3)&$ 3d_{5/2} -  5f_{5/2}$&    1.315(6)\\
$ 3d_{3/2} -  6p_{3/2}$&    0.464(5) & $ 3d_{3/2} -  6f_{5/2}$&    2.899(8) & $ 3d_{5/2} -  6f_{7/2}$&    3.465(0)&$ 3d_{5/2} -  6f_{5/2}$&    0.775(2)\\
$ 3d_{3/2} -  7p_{3/2}$&    0.224(2) & $ 3d_{3/2} -  7f_{5/2}$&    2.001(5) & $ 3d_{5/2} -  7f_{7/2}$&    2.392(6)&$ 3d_{5/2} -  7f_{5/2}$&    0.535(1)\\
$ 3d_{3/2} -  8p_{3/2}$&    0.143(1) &                        &             &                        &            &                       & \\
$ 3d_{3/2} -  9p_{3/2}$&    0.103(1) && &&  & &\\
$ 3d_{3/2} - 10p_{3/2}$&    0.079(1) && &&  & &\\
\end{tabular}
\end{ruledtabular}
\end{table*}

  Magic wavelengths for
the alkali-metal atoms from Na to Cs, for which the $ns$ and $np_{1/2}$ or $np_{3/2}$ atomic levels have the same ac
Stark shifts, were evaluated by Arora {\it et al.\/}
\cite{AroSafCla07}. The case of circular polarization  was considered in~\cite{AroSah12,SahAro12}.  McKeever et al. \cite{MckBucBoo03} demonstrated state-insensitive
trapping of Cs atoms at 935~nm while maintaining a strong coupling for the $6s_{1/2}-6p_{3/2}$ transition. A
bichromatic scheme for state-insensitive optical trapping of Rb atom was explored in Ref.~\cite{AroSafCla10}. In the
case of Rb, the magic wavelengths associated with monochromatic trapping were sparse and relatively inconvenient. The
bichromatic approach yielded a number of promising magic wavelength pairs. The precise
  magic wavelengths for Li $2s-2p_j$ and $2s-3p_j$ transitions in convenient
  wavelength regions were recently calculated
  in~\cite{Li}.  The results were presented for both $^6$Li and $^7$Li
   to illustrate the possibilities
  for differential light shifts between the two isotopes.

\subsection{Group II}
\label{secaII}

The magic wavelengths for the Sr $5s^2~^1S_0 - 5s5p\ ^3P^{\circ}_0$ and $5s^2~^1S_0 - 5s5p\ ^3P^{\circ}_1$ transitions have been
measured in \cite{Sr,Sr1}. The Yb clock $6s^2~^1S_0 - 6s6p\ ^3P^{\circ}_0$ magic wavelength was predicted to be 752~nm in
\cite{Der} and measured to be 759.355~nm in Ref.~\cite{Yb}. The magic wavelength for the ultraviolet $6s^2\ ^1S_0
\leftrightarrow 6s6p\ ^3P^{\circ}_0$ clock transition in Hg was recently reported by Yi {\it et al.\/} \cite{YiMejMcF11}.
 The Stark-free (magic) wavelength was found to be 362.53(0.21)~nm, in excellent
  agreement with the theoretical prediction
 360~nm from ~\cite{Hg1}, calculated using a method that combines
 configuration interaction and many-body perturbation theory.
The magic wavelengths of other group II and group IIb atoms have been predicted in \cite{DerObrDzu09}.
 The magic wavelengths
  are very sensitive to the values of the ac polarizabilities and
  allow for precise tests of the theory ~\cite{usYb,usSr}. Moreover, the magic
wavelengths can be used to determine the values of
important electric-dipole matrix elements which are difficult to obtain
  by direct experimental techniques.
  For example, the $5s5p~^3P^{\circ}_0 - 5s6s~^3S_1$ matrix element in Sr was recently determined
  using the experimental value of
  the Sr $5s^2~^1S_0 - 5s5p\ ^3P^{\circ}_0$ magic wavelength with 0.5\% precision ~\cite{usSr}.
Dammalapati {\it et al.\/} \cite{DamSWil12} investigated light shifts
of heavy
 alkaline earth elements barium (Ba) and
radium (Ra), which are of interest  for development of optical
lattice clocks
 and for permanent electric dipole moment
searches. The wavelength dependence of light shifts of the $ns^2\
^1S_0$ ground state,
 the $nsnp\ ^3P^{\circ}_1$ and $ns(n-1)d\ ^1D_2$ excited states in Ba ($n$ = 6) and the $ns^2\ ^1S_0$ ground state, the $nsnp\ ^3P^{\circ}_1$ and $ns(n - 1)d\
^3D_2$ excited states in Ra ($n$ = 7) were calculated. Several
magic wavelengths in the visible and infrared regions accessible with
commercial lasers for optical dipole trapping of Ba and Ra were
identified \cite{DamSWil12}. Magic wavelengths of an optical clock
transition of barium were presented in \cite{GenJiaRun11}. Dipole
polarizabilities of $ns^2\ ^1S_0$ and $nsnp\ ^3P^{\circ}_0$ states and
relevant magic wavelengths of Sr, Yb, Zn, Cd, and Hg atoms were
studied by a semiempirical approach in
Refs.~\cite{YeWan08,GuoWanYe10}.

The magic wavelength conditions that can make optical lattice clocks
insensitive to atomic motion were presented by Katori {\it et al.\/}
\cite{KatHasIli09}. This work demonstrated that the spatial mismatch
of the interactions in the clock transition can be treated as a
spatially constant offset $\delta\nu$  for specific lattice
geometries. Numerical estimates were made for Sr \cite{KatHasIli09}.

Theoretical study of the dynamic scalar polarizabilities of the
ground and selected long-lived excited states of dysprosium was
recently carried out by Dzuba {\it et al.\/} \cite{DzuFlaLev11}. A
set of the magic wavelengths of the unpolarized lattice laser field
for each pair of states, which includes the ground state and one of
these excited states was given. The authors presented an analytical
formula that allows for the determination of approximate values of
the magic wavelengths without calculating the dynamic
polarizabilities of the excited states near resonances
\cite{DzuFlaLev11}.

\begin{table}
\caption{\label{tab-4p-com} Values of scalar ($\alpha_{0}$)
 and tensor ($\alpha_{2}$)  polarizabilities in K.
 The present results are compared with  theoretical and experimental   values. Ref.~\cite{SafSaf08}
 did not include uncertainty estimates.
 All values are in atomic units.}
\begin{ruledtabular}
\begin{tabular}{lllll}
\multicolumn{1}{c}{}& \multicolumn{1}{c}{Present}& \multicolumn{1}{c}{Theory}&
\multicolumn{1}{c}{Expt.}\\[0.2pc]\hline
$\alpha_{0}(4s_{1/2})$& 290.4(6)& 290.2(8)\cite{DerJohSaf99} &  290.8(1.4)\cite{HolRevLon10}  \\
$\alpha_{0}(5s_{1/2})$& 4961(22)&  &    \\[0.4pc]
$\alpha_{0}(4p_{1/2})$&      611(6)& 604.1 \cite{SafSaf08}  &  587(87) \cite{MarYel69}\\
$\alpha_{0}(4p_{3/2})$&      620(5)& 614.1\cite{SafSaf08}   &  613(103) \cite{MarYel69}\\
$\alpha_{0}(5p_{1/2})$& 7053(70)&  &    \\
$\alpha_{0}(5p_{3/2})$& 7230(61)&  &    \\[0.4pc]
$\alpha_{0}(3d_{3/2})$& 1420(30)&  &    \\
$\alpha_{0}(3d_{5/2})$& 1412(31)&  &    \\[0.4pc]
$\alpha_{2}(4p_{3/2})$& -109.4(1.1)& -107.9\cite{SafSaf08}  &                       \\
$\alpha_{2}(5p_{3/2})$& -1065(18)&  &    \\[0.4pc]
$\alpha_{2}(3d_{3/2})$& -482(19)&  &    \\
$\alpha_{2}(3d_{5/2})$& -673(23)&  &    \\
\end{tabular}
  \end{ruledtabular}
\end{table}

\section{Matrix elements and polarizabilities}
\label{sec1}
 The magic wavelengths for a specific transition are located by calculating the
 frequency-dependent
 polarizabilities of the
lower and upper states and finding their crossing points. The
all-order approach to the calculation of atomic polarizabilities was
discussed  in
Refs.~\cite{AroSafCla07,JonSafDer08,SafSaf08,AroSafCla11,SafSaf11,Li},
and we provide only a brief summary of the methods here. Unless
stated otherwise, all specific data refers to the K atom, and we use the conventional system of atomic units,
a.u., in which $e, m_{\rm e}$, $4\pi \epsilon_0$ and the reduced
Planck constant $\hbar$ have the numerical value 1.  Polarizability
in a.u. has the dimension of volume, and its numerical values
presented here are expressed in units of $a^3_0$, where
$a_0\approx0.052918$~nm is the Bohr radius. The atomic units for
$\alpha$ can be converted to SI units via
 $\alpha/h$~[Hz/(V/m)$^2$]=2.48832$\times10^{-8}\alpha$~[a.u.], where
 the conversion coefficient is $4\pi \epsilon_0 a^3_0/h$ and the
 Planck constant $h$ is factored out.

 The
frequency-dependent scalar polarizability, $\alpha(\omega)$, of an
alkali-metal atom in the state $v$ may be separated into a
contribution from the ionic core, $\alpha_{\rm{core}}$, a core
polarizability modification due to the valence electron,
$\alpha_{vc}$, and a contribution from the valence electron,
$\alpha^v(\omega)$.
 The valence contribution to frequency-dependent scalar $\alpha_0$ and
 tensor $\alpha_2$ polarizabilities is
evaluated as the sum over intermediate $k$ states allowed by the electric-dipole transition rules~\cite{MitSafCla10}
\begin{eqnarray}
    \alpha_{0}^v(\omega)&=&\frac{2}{3(2j_v+1)}\sum_k\frac{{\left\langle k\left\|d\right\|v\right\rangle}^2(E_k-E_v)}{     (E_k-E_v)^2-\omega^2}, \label{eq-1} \nonumber \\
    \alpha_{2}^v(\omega)&=&-4C\sum_k(-1)^{j_v+j_k+1}
            \left\{
                    \begin{array}{ccc}
                    j_v & 1 & j_k \\
                    1 & j_v & 2 \\
                    \end{array}
            \right\} \nonumber \\
      & &\times \frac{{\left\langle
            k\left\|d\right\|v\right\rangle}^2(E_k-E_v)}{
            (E_k-E_v)^2-\omega^2} \label{eq-pol},
\end{eqnarray}
             where $C$ is given by
\begin{equation}
            C =
                \left(\frac{5j_v(2j_v-1)}{6(j_v+1)(2j_v+1)(2j_v+3)}\right)^{1/2} \nonumber
\end{equation}
and ${\left\langle k\left\|d\right\|v\right\rangle}$ are the reduced electric-dipole matrix elements. In these
equations, $\omega$ is assumed to be at least several linewidths off resonance with the corresponding transitions.
Linear polarization is assumed in all calculation. The ionic core polarizability and $\alpha_{vc}$ term depend weakly
on $\omega$ for the frequencies treated here and are approximated by their dc values calculated in the random-phase
approximation (RPA) ~\cite{MitSafCla10}. We find the contribution from the K$^+$ ionic core to be $\alpha
_{\text{core}}= 5.457~a_0^{3}$. A counter term $\alpha _{\rm vc}$ compensating for excitation from the core to the
valence shell which violates the Pauli principle is very small. For example, it is  $\alpha _{\rm vc} = -0.00015$~a.u.
for the $4p_j$ states of K.

We use the linearized version of the coupled cluster approach
  (also referred to as the all-order method), which sums infinite sets
of many-body perturbation theory terms, for all significant terms in
the equations above. The  $4s-np$, $4p-nl$, $5s-nl$, $5p-nl$, and
$3d-nl$ transitions with $n\leq 26$ are calculated
 using this approach~\cite{SafJoh08,SafSaf11}.
\begin{table}
\caption[]{Magic wavelengths for the $4s-np_j$ transitions in K.
 The $500-1227$~nm and $1050-1130$~wavelength ranges
were considered for the $4s-4p_j$ and $4s-5p_j$ transitions, respectively. The corresponding polarizabilities are given
in a.u. The resonance near the magic wavelengths are listed in the first column.} \label{tab3}
\begin{ruledtabular}
\begin{tabular}{lrr}
   \multicolumn{1}{c}{Resonance} &
   \multicolumn{1}{c}{K $ \lambda_\textrm{magic}$} &
  \multicolumn{1}{c}{$\alpha$} \\
\hline
   \multicolumn{3}{c}{$4s-4p_{1/2}$ Transition}\\
  $4p_{1/2} - 9s      $            &  508.12(1)     &  -215(2)\\
  $4p_{1/2} - 7d_{3/2}$            &  509.47(1)    &  -220(3)\\
  $4p_{1/2} - 8s      $            &  531.80(1)    &  -256(2)\\
  $4p_{1/2} - 6d_{3/2}$            &  533.99(1)    &  -260(3)\\
  $4p_{1/2} - 7s      $            &  577.37(1)    &  -365(3)\\
  $4p_{1/2} - 5d_{3/2}$            &  581.05(1)    &  -375(4)\\
  $4p_{1/2} - 6s      $            &  690.17(1)    &  -1195(10)\\
  $4p_{1/2} - 4s      $            &  768.41(1)    &  21200(400)\\
  $4p_{1/2} - 5s,3d_{3/2}$   &  1227.63(2)    &  475(40)\\[0.2pc]
   \multicolumn{3}{c}{$4s-4p_{3/2}$, $|m_j|=1/2$ Transition}\\
  $4p_{3/2} - 9s      $ &  509.38(1)  &  -217(2)\\
  $4p_{3/2} - 7d_{5/2}$ &  511.04(1)  &  -220(2)\\
  $4p_{3/2} - 8s      $ &  533.07(1)  &  -259(2)\\
  $4p_{3/2} - 6d_{5/2}$ &  535.72(1)  &  -264(2)\\
  $4p_{3/2} - 7s      $ &  578.71(1)  &  -369(3)\\
  $4p_{3/2} - 5d_{5/2}$ &  583.07(1)  &  -383(3)\\
  $4p_{3/2} - 6s      $&  692.35(1)  &  -1237(12)\\
  $4p_{3/2} - 4s      $&  769.43(1)  & -27400(200)\\
  $4p_{3/2} - 5s,3d_{j}$&  1227.61(1) &  475(45)\\[0.2pc]
    \multicolumn{3}{c}{$4s-4p_{3/2}$, $|m_j|=3/2$ Transition}\\
   $4p_{3/2} - 7d_{5/2}$ &   510.75(1)   & -319(2)   \\
   $4p_{3/2} - 6d_{5/2}$ &   535.38(1)   & -263(3)   \\
   $4p_{3/2} - 5d_{5/2}$ &   582.80(1)   & -383(3)   \\
   $4s-4p_{j}          $ &  768.98(2)   & -367\\[0.2pc]
   \multicolumn{3}{c}{$4s-5p_{1/2}$ Transition}\\
  $5p_{1/2} - 14s      $ &1050.238(2)  & 620(5)  \\
  $5p_{1/2} - 12d_{3/2}$ &1051.528(2)  & 620(5) \\
  $5p_{1/2} - 13s      $ &1067.326(2)  & 600(5) \\
  $5p_{1/2} - 11d_{3/2}$ &1069.017(2)  & 598(5) \\
  $5p_{1/2} - 12s      $ &1090.784(4)  & 574(5) \\
  $5p_{1/2} - 10d_{3/2}$ &1093.057(3)  & 572(5) \\
  $5p_{1/2} - 11s      $ &1124.419(6)  & 543(5) \\
  $5p_{1/2} - 9d_{3/2} $ &1127.560(6)  & 540(5) \\[0.2pc]
   \multicolumn{3}{c}{$4s-5p_{3/2}$, $|m_j|=1/2$ Transition}\\
  $5p_{3/2} - 14s      $  &1052.438(2)   & 618(5) \\
  $5p_{3/2} - 12d_{3/2}$  &1053.647(2)   & 617(5) \\
  $5p_{3/2} - 13s      $  &1069.656(3)   & 597(5) \\
  $5p_{3/2} - 11d_{3/2}$  &1071.224(3)   & 595(5) \\
  $5p_{3/2} - 12s      $  &1093.312(4)   & 571(5) \\
  $5p_{3/2} - 10d_{3/2}$  &1095.393(4)   & 569(5) \\
  $5p_{3/2} - 11s      $  &1127.275(8)   & 540(5)\\
  $5p_{3/2} - 9d_{3/2} $  &1130.092(8)   & 538(5)\\    [0.2pc]
   \multicolumn{3}{c}{$4s-5p_{3/2}$, $|m_j|=3/2$ Transition}\\
 $5p_{3/2} - 12d_{3/2}$   &1053.593(2)   & 617(5)    \\
 $5p_{3/2} - 11d_{3/2}$   &1071.144(2)   & 595(5)\\
 $5p_{3/2} - 10d_{3/2}$   &1095.274(3)   & 569(5)\\
 $5p_{3/2} -  9d_{3/2}$   &1129.908(4)   & 538(5)\\
 \end{tabular}
\end{ruledtabular}
\end{table}

As we noted in the Introduction, the present calculation required
evaluation of the electric-dipole matrix elements for highly excited
states, since the frequency-dependent polarizabilities for the
$4s-5p$ magic wavelengths of particular experimental interest are
dominated by the $5p-nl$ transitions with $n=12-14$. The difficulty
with the applications of the all-order method for these states
results from the use a complete set of Dirac-Fock (DF) wave functions
on a nonlinear grid generated using B-splines constrained to a
spherical cavity. A large cavity  with radius of $R = 220~a_0$ is
needed to accommodate all valence orbitals with $ns=4s-10s$,
$np=4p-10p$, and $nd=3d-9d$. A cavity radius of 400~$a_0$ was chosen
to accommodate additional valence orbitals with $ns=11s-14s$,
$np=11p-13p$, and $nd=10d-12d$. Our basis set consists of 70 splines
of order 11 for each value of the relativistic angular quantum number
$\kappa$ for $R = 220~a_0$ and  100 splines of order 13 for $R =
400~a_0$. We have conducted test comparisons of the basis set
energies with the actual DF values to demonstrate the numerical
stability of this calculation.
\begin{figure*}[tbp]
\includegraphics[scale=0.8]{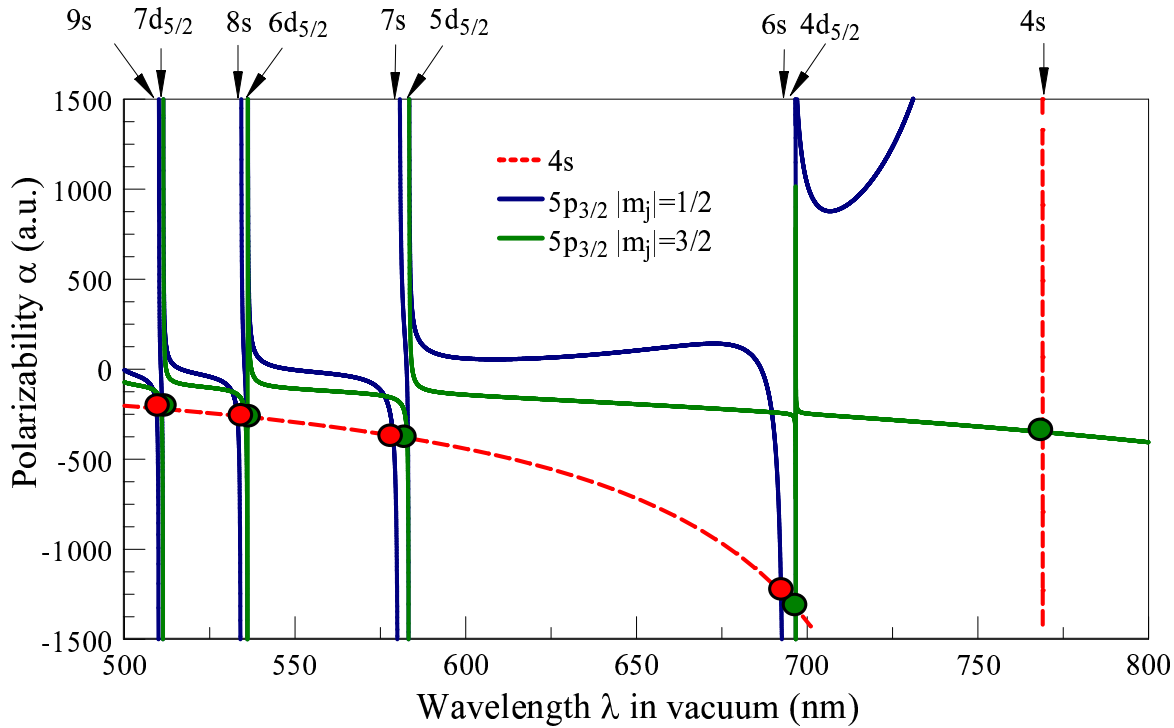}
\caption{(Color online) The frequency-dependent polarizabilities of the K $4s$ and $4p_{3/2}$ states.
  The magic wavelengths are marked with circles. The approximate positions of the $4p_{3/2}-nl$ resonances are indicated by
vertical lines with small arrows on top of the graph, together with the corresponding $nl$.} \label{fig1}
\end{figure*}
We can use available experimental energies for the $ns=4s-11s$, $np=4p-10p$, and $nd=3d-12d$ states from
\cite{RalKraRea11}  and theoretical all-order energies for other states with $n\leq26$. The remaining small
contributions with $n>26$ are calculated in the DF approximation. For example, the contributions from states with
$n>26$ give only 0.075~a.u. to the polarizability of the $4p_{1/2}$ state. We note that states with  $n>19$ in our
basis have positive energies and provide a discrete representation of the continuum.

 The evaluation of the uncertainty of the matrix elements in this approach was
described in detail in \cite{SafSaf11,Li}. Four all-order
calculations were
 carried out, including two  \textit{ab initio} all-order calculations with
and without the inclusion of the partial triple excitations and two
 other calculations that incorporated semiempirical estimates
of high-order correlation corrections starting from both \textit{ab initio} runs.
The spread of these
 four values for each transition defines the  estimated uncertainty in the final
results when considered justified based on the dominant correlation
contributions to the E1 matrix elements~\cite{SafSaf11,Li}. We note that this procedure 
does not work in the small number of cases where we can not estimate uncertainty in the dominant contributions
using the procedure described above. No uncertainties are listed for such matrix elements, however,
 their contributions are small, leading to negligible effects on the final uncertainties of the polarizabilities.
The
absolute values of the reduced electric-dipole matrix elements
 used in our subsequent calculations and their uncertainties are listed in a.u.
in Table~\ref{tab1}. We list only the  most important subset of the several hundred matrix elements that were
calculated in this work.

Our results for scalar and tensor polarizabilities  of the $4p_{j}$ excited
states of potassium  are compared  with
calculations of \cite{SchLurHap71,MerBeg98} and with experimental measurements
reported by \citet{MarYel69} in
Table~\ref{tab-4p-com}. The Bates-Damgaard method was used by
 \citet{SchLurHap71} and the time-dependent
gauge-invariant variational method was used by \citet{MarYel69}.
 The uncertainty in the experimental
measurement~\cite{MarYel69} of the scalar polarizability is too large to
 reflect on the accuracy of the present
calculations. Extensive comparison of the theoretical and experimental
 static polarizabilities for the alkali-metal atoms was recently given
 in the review ~\cite{MitSafCla10}.

\section{Magic wavelengths}
\label{sec3}

\begin{figure}[tbp]
\includegraphics[width=2.7in]{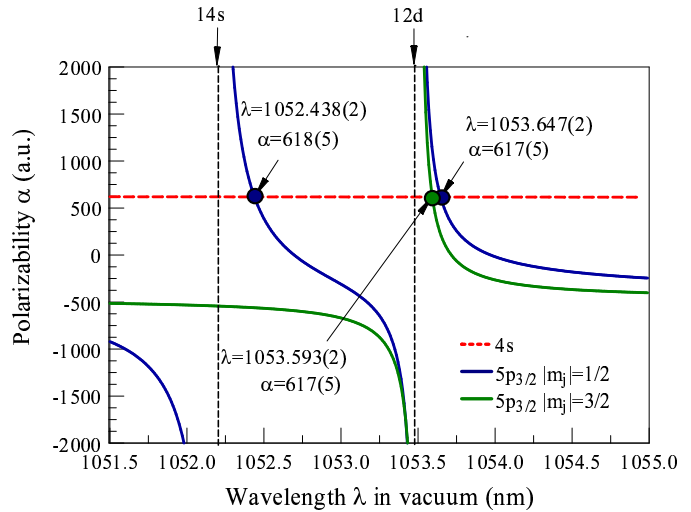}
                         \caption{(Color online) The frequency-dependent polarizabilities of the K $4s$ and $5p_{3/2}$ states.
  The magic wavelengths are marked with circles and arrows. The approximate positions of the $5p_{3/2}-14s$  and $5p_{3/2}-12d$ resonances are indicated by
vertical lines with small arrows on top of the graph.} \label{fig2}
\end{figure}

\begin{figure*}[tbp]
\includegraphics[scale=0.7]{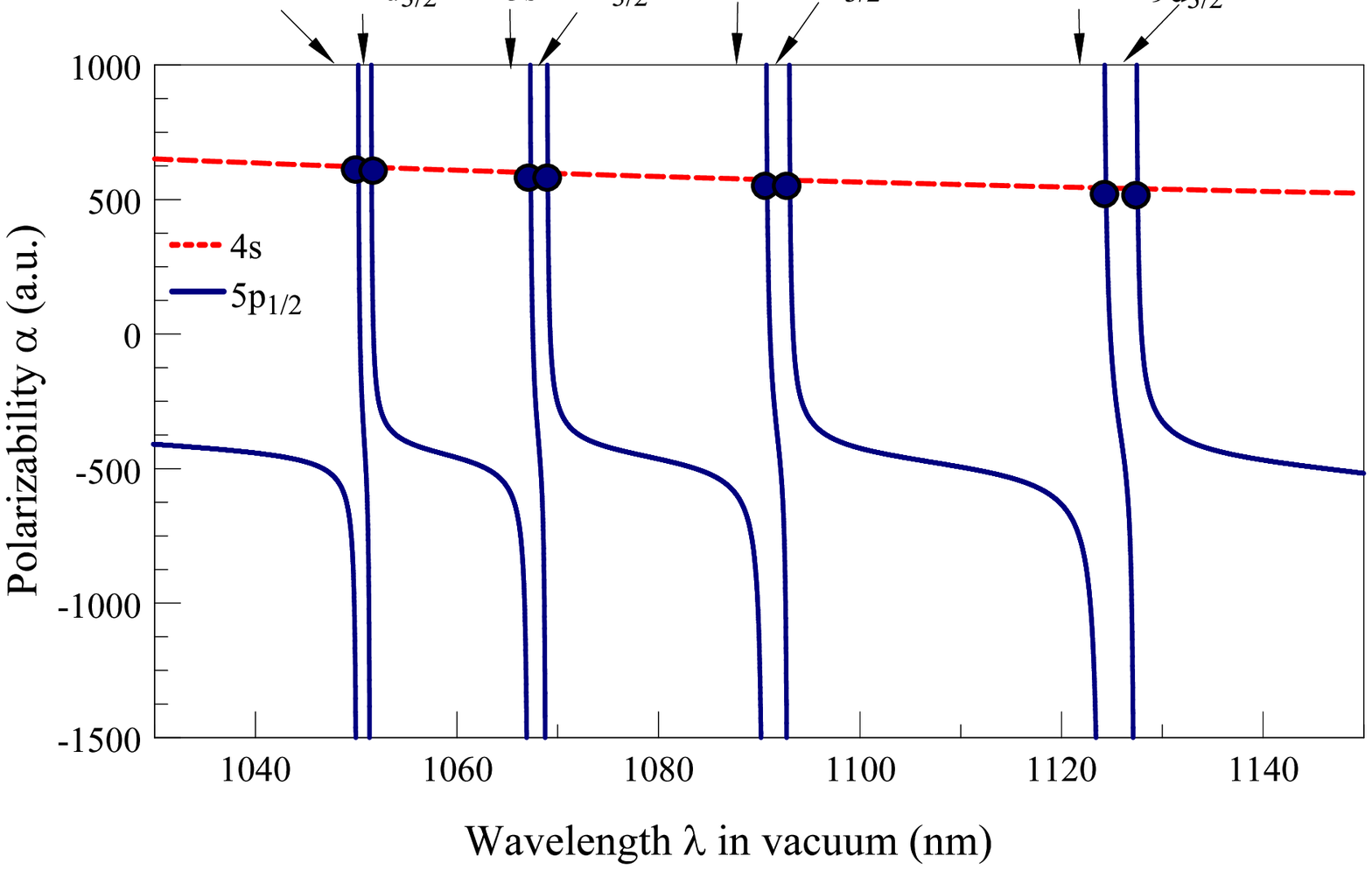}
\caption{(Color online) The frequency-dependent polarizabilities of the K $4s$ and $5p_{1/2}$ states.
  The magic wavelengths are marked with circles. The approximate positions of the $5p_{1/2}-nl$ resonances are indicated by
vertical lines with small arrows on top of the graph, together with the corresponding $nl$.} \label{fig3}
\end{figure*}

\begin{figure*}[tbp]
\includegraphics[scale=0.7]{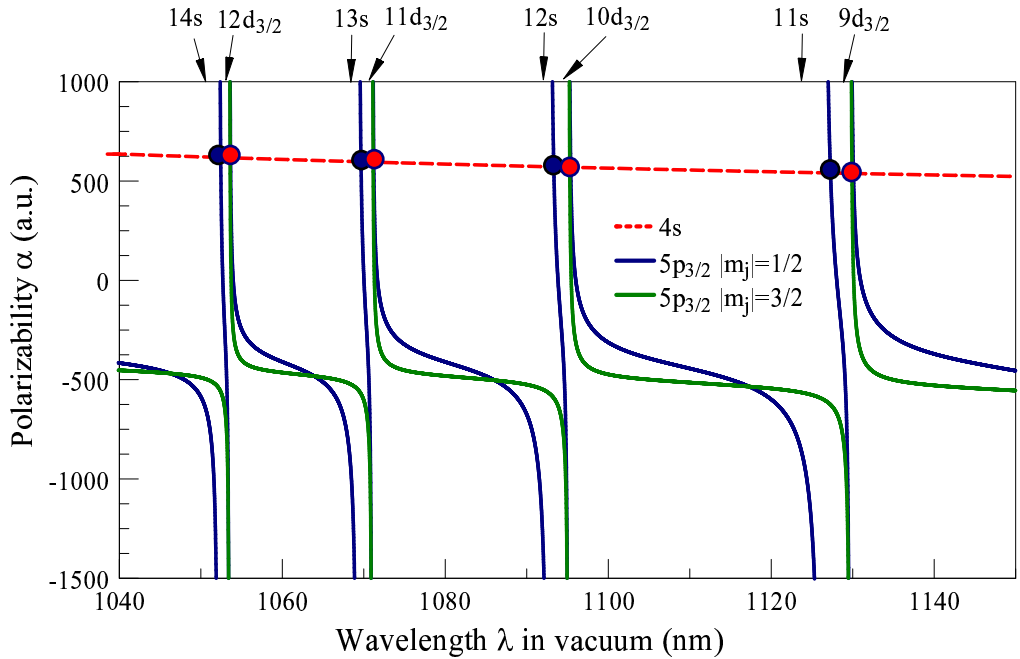}
\caption{(Color online) The frequency-dependent polarizabilities of the K $4s$ and $5p_{3/2}$ states.
  The magic wavelengths are marked with circles. The approximate positions of the $5p_{3/2}-nl$ resonances are indicated by
vertical lines with small arrows on top of the graph, together with the corresponding $nl$.} \label{fig4}
\end{figure*}

We define the magic wavelength $\lambda_{\rm{magic}}$ as the wavelength
 for which the ac polarizabilities of  two states involved in the atomic transition
  are the same, leading to a
vanishing ac Stark shift of that transition. For the $ns-np$ transitions,
a magic wavelength is represented by the
point at which two curves, $\alpha_{ns}(\lambda)$ and $\alpha_{np}(\lambda)$,
intersect as a function of the wavelength
$\lambda$. The total polarizability for the $np_{3/2}$ states is given by
 $\alpha=\alpha_0-\alpha_2$ for $m_j=\pm 1/2$
and $\alpha=\alpha_0+\alpha_2$  for the $m_j=\pm 3/2$ case. Therefore,
the total polarizability of the $np_{3/2}$
state depends
 upon its $m_j$ quantum number and
  the magic wavelengths need to be determined separately for the cases
  with $m_j=\pm 1/2$ and
$m_j=\pm 3/2$  for the $ns-np_{3/2}$ transitions, owing to the presence of
 the tensor contribution to the total
polarizability of the $np_{3/2}$ state.
 The uncertainties in the values of magic wavelengths are found as the maximum
 differences between
 the central value and the crossings of the
$\alpha_{ns} \pm \delta \alpha_{ns}$ and $\alpha_{np} \pm \delta \alpha_{np}$
 curves, where the  $\delta \alpha$
are the uncertainties in the corresponding $ns$ and $np$ polarizability values.
 All calculations are carried out for linear polarization.
Several magic wavelengths were calculated for the $4s-4p_{1/2}$ and $4s-4p_{3/2}$ transitions in K in
Ref.~\cite{AroSafCla07} using the all-order approach. Only the magic wavelengths with $\lambda>600$~nm
 were listed. In this work, we present
several other magic wavelengths for these $D_1$, $D_2$ transitions above 500~nm.

The frequency-dependent polarizabilities of the $4s$ and $4p_{3/2}$
states for $\lambda=500 - 800$~nm are plotted in  Fig.~\ref{fig1}.
The magic wavelengths are marked with circles. The approximate
positions of the $4p_{3/2}-nl$ resonances are indicated by vertical
lines with small arrows on top of the graph, together with the
corresponding $nl$. For example, the arrow labelled $7s$ indicates
the position of the $4p_{3/2}-7s$ resonance.
 The corresponding magic wavelengths are listed in Table~\ref{tab3}.
 We note that the $4p_{3/2}-5s$
resonance wavelength is outside of the plot region at $\lambda=1253$~nm).
   While there are 8 magic wavelengths for the
$4s-4p_{3/2} ~|m_j|=1/2$ transition in the wavelength region shown on the plot,
 there are only 4 magic wavelengths for the $4s-4p_{3/2} ~|m_j|=3/2$
 transition
  since there are no corresponding
crossings near the $4p_{3/2}-ns$ resonances as in the case of
$|m_j|=1/2$. The 769~nm magic wavelength for the $|m_j|=1/2$ is not
shown on the plot since the corresponding polarizability
(-27400~a.u.)
 is outside of the plot y-axis range.
There is only one magic wavelength above 800~nm for the $4s-4p_{3/2} ~|m_j|=1/2$
transition
due to $4p_{3/2}-5s$ resonance and
none for the $|m_{j}|=3/2$ case.  The magic wavelengths for the $4s-4p_{1/2}$
 transition are very close to those for
$4s-4p_{3/2} ~|m_j|=1/2$. They are also given in Table~\ref{tab3}.

The magic wavelengths for the UV $4s-5p_j$ transitions are completely
different than those for the $D_1$, $D_2$
lines owing to completely different set of resonances. The K case is also
 significantly different from that of Li \cite{Li} due to differences in the resonant
 transition wavelengths. We list the
magic wavelengths for the $4s-5p_{1/2}$ and $4s-5p_{3/2}$ transitions in the range of 1050-1130~nm, which is of
particular experimental interest in Table~\ref{tab3}.
We find 20 magic wavelengths in the technically interest region of $1050-1130$~nm accessible by a number of widely used lasers. 
 The magic wavelengths for
 the $4s-5p_{3/2}$ transition  near
1053~nm wavelength  are illustrated in Fig.~\ref{fig2}. As in the case of the
 $4s-4p_{3/2}$ transition, there is no
magic wavelength for the $|m_j|=1/2$ case near the $ns$ resonance.
All magic wavelengths for the $4s-5p_{1/2}$ and
$4s-5p_{3/2}$ transitions in the range of 1050-1140~nm  are illustrated
in Figs.~\ref{fig3} and Figs.~\ref{fig4}. The
same designations are used as in the previous graphs.
 Comparing these figures with the similar plots for Li (see
Figs.~3 and 4 of Ref.~\cite{Li}) shows that K magic wavelengths near 1050-1130~nm originate from crossings near much
higher resonances ($n=9-14$ vs. $n=6-7$ for Li) making the calculation for K
 more complicated due to the very large cavity size
required to accommodate such highly excited orbitals.

\section{Conclusion}
We have calculated the ground $4s$, $4p$, and $5p$ state ac polarizabilities
 in K using the relativistic linearized
coupled-cluster method and evaluated the uncertainties of these values.
 We have used our calculations
 to identify the magic wavelengths for the $4s-4p$ and $4s-5p$
 transitions. The  magic wavelengths for the  ultraviolet resonance
  lines is of particular interest for
 laser cooling of ultracold gases with high phase-space densities.

\section*{Acknowledgement}
This research was performed under the sponsorship of the US Department of Commerce, National Institute of Standards and Technology, and was supported
by the National Science Foundation under Physics Frontiers Center Grant PHY-0822671.


\end{document}